\documentclass[a4paper,11pt]{article}
\usepackage[utf8]{inputenc}
\pdfoutput=1
\usepackage{graphicx}
\usepackage{amsmath,amssymb,mathrsfs}
\usepackage{bbm}
\usepackage{color}
\usepackage{xcolor}
\usepackage{dsfont} 
\usepackage{cancel}
\usepackage{dsfont}
\usepackage{epstopdf}
\usepackage{epsfig}
\usepackage{bm}
\usepackage{authblk}
\usepackage{dcolumn}
\usepackage{enumitem}
\usepackage{multirow}
\usepackage{lineno}
\usepackage{mathtools}
\usepackage{url}
\usepackage{lineno}
\usepackage{wrapfig}
\usepackage{caption}
\usepackage{subcaption}
\usepackage{jheppub}
\usepackage[toc,page]{appendix}
\usepackage{wrapfig}
\usepackage[T1]{fontenc}

\makeatletter
\gdef\@fpheader{}
\g@addto@macro\bfseries{\boldmath}
\makeatother

\title{Steady-state bubbles beyond local thermal equilibrium}
\author[1]{Tomasz~Krajewski,}
\author[2]{Marek~Lewicki,}
\author[2]{Ignacy~Nałęcz}
\author[2]{and Mateusz~Zych}
\affiliation[1]{Nicolaus Copernicus Astronomical Center of the  Polish Academy of Sciences, ul. Bartycka 18, 00-716 Warsaw, Poland}
\affiliation[2]{Faculty of Physics, University of Warsaw, ul.\ Pasteura 5, 02-093 Warsaw, Poland}

\emailAdd{tkrajewski@camk.edu.pl}
\emailAdd{marek.lewicki@fuw.edu.pl}
\emailAdd{ignacy.nalecz@fuw.edu.pl}
\emailAdd{mateusz.zych@fuw.edu.pl}

\abstract{
We investigate the hydrodynamic solutions for expanding bubbles in cosmological first-order phase transitions going beyond local thermal equilibrium approximation. Under the assumption of a tangensoidal field profile, we supplement the matching conditions with the entropy produced due to the interaction of the bubble wall with ambient plasma. This allows us to analytically compute the corresponding fluid profiles and find the bubble-wall velocity. We show that due to the entropy production, two stable solutions corresponding to a deflagration or hybrid and a detonation can coexist. Finally, we use numerical real-time simulations of bubble growth to show that in such cases it is typically the faster detonation solution which is realised. This effect can be explained in terms of the fluid profile not being fully formed into the predicted steady-state solution as the wall accelerates past this slower solution.
}
\begin{document}

\maketitle
\section{Introduction}

Cosmological phase transitions are fascinating processes occurring in the early stages of the Universe evolution. Their rich phenomenology can provide solutions to some of the greatest puzzles in modern cosmology. First order phase transitions (FOPTs) lead to departure from thermal equilibrium facilitating the rare environment in which Sakharov's conditions are satisfied. Electroweak baryogenesis~\cite{Kuzmin:1985mm,Cohen:1993nk,Rubakov:1996vz,Morrissey:2012db} realized at the surfaces of growing bubbles, nucleated during the transition, can explain the origin of the observable baryon asymmetry. Furthermore, when the bubble walls collide, their kinetic energy is released and transformed into gravitational waves, producing a stochastic background which might be detected in the near future~\cite{Caprini:2015zlo,Caprini:2019egz,Badurina:2021rgt,LISACosmologyWorkingGroup:2022jok,Caprini:2024hue,Caprini:2024ofd}.

From the theoretical point of view, cosmological scalar phase transitions are a common theme in many particle physics models. Generally speaking, interactions with the hot primordial plasma drive the symmetry-breaking field into symmetry-preserving vacuum states. As the Universe cools down, the impact of the plasma diminishes, and eventually, at the critical temperature $T_c$, the symmetry-breaking and symmetry-preserving vacua become degenerate. Below this temperature, the scalar field evolves toward the symmetry-breaking configuration unless it is forbidden by the energy barrier that prevents the field from rolling toward the energetically favourable state. Even if a smooth transition to the new phase is not possible, the field may tunnel into it quantum mechanically or via thermal fluctuations, forming the small spherical regions (bubbles) of the new phase. The temperature $T_n$ of the background plasma at which this process becomes efficient, i.e., when the probability of nucleating a single bubble within a Hubble horizon is of the order~1, is called the nucleation temperature. Once nucleated, bubbles start to expand, driven by the energy difference between the supercooled and true vacua. In some cases, the driving force is balanced by field interactions with the primordial plasma, which prevent the bubble walls from accelerating beyond a terminal velocity $v_w$.

Precise computation of the terminal wall velocity is crucial for accurate predictions of the stochastic gravitational wave signal from FOPTs and faithful estimation of the baryogenesis yield. Despite its importance, the correct computation of this parameter remains a challenge. It was recently shown that in the local thermal equilibrium (LTE) approximation, it is possible to determine the bubble wall velocity in a steady state in a relatively simple way~\cite{BarrosoMancha:2020fay, Ai:2021kak, Ai:2023see}. The method is based on the conservation of entropy and provides the additional matching equation relating its value across the bubble wall. Since additional friction can only slow down the wall, the resulting velocity can be interpreted as the upper bound on the full out-of-equilibrium computation. It was also demonstrated, using the example of a scalar singlet extension of the Standard Model, that non-equilibrium corrections are typically subdominant~\cite{Laurent:2022jrs}.

Nevertheless, this extremely useful method, similarly to all steady-state results does not take into account the dynamical formation of the heated fluid shell, which is the source of hydrodynamic obstruction. In the recent paper~\cite{Krajewski:2024gma} we have shown that majority of deflagration and hybrid solutions predicted by the LTE matching equations are not realized in dynamical simulations and that real-time evolution towards deflagration/hybrid requires fine-tuning of the nucleation temperature $T_n$ to be very close to the critical temperature $T_n \approx T_c$, i.e. the one at which initial and final vacua are degenerate.

In this paper we generalize the matching method, going beyond the LTE approximation. We obtain a new version of the third matching condition by introducing the entropy production rate.We assume a particular class of interactions between the wall and the plasma~\cite{Moore:1995si,John:2000zq}, %valid for the SM and many of its extensions~\cite{Moore:1995si,John:2000zq}, 
expressing the corresponding interaction rate in the thin wall regime using basic physical quantities and the effective, phenomenological friction parameter~\cite{Ignatius:1993qn}.

The rate at which entropy is generated in the field-plasma interaction critically depends on the stationary field profile during the transition and in particular its width. Surprisingly, in the hydrodynamic lattice simulations, we find this width to be independent of friction, which allows us to estimate it using only the information stored in the effective potential. The resulting analytical field profiles are then used to compute the entropy source necessary to obtain the steady-state fluid profile in the corresponding model.

Intriguingly, our new formalism typically predicts an additional branch of stable detonation solutions. This branch of solutions is well visible in numerical simulations, continuously going to runaways with $v_w = 1$ in the limit of vanishing entropy production rate. The presence of this branch gives a convenient selection rule to distinguish between physical and unphysical deflagration/hybrid solutions as it is typically the faster detonation solution which is dynamically realized. This is associated with the fact that the steady state profile predicted analytically is not fully formed as the developing solution accelerates past the velocity corresponding to the slower solution~\cite{Krajewski:2024gma}.

We supplement our paper with a code snippet written in \texttt{Python} that can determine the stationary wall velocity for both detonations and deflagrations/hybrids beyond LTE~\cite{Repo:2024in}. Our code is essentially a modified version of the program published in~\cite{Ai:2023see}. The wall velocity is computed based on parameter encoding dissipative friction and the parameters used for calculating $v_w$ in LTE that are readily derived from the effective potential. We tested our algorithm by comparing its predictions with the results of hydrodynamic lattice simulations~\cite{Krajewski:2023clt} and found that it correctly reproduces the stationary velocity to great accuracy.

The paper is organized as follows. In Section \ref{sec:2}, we recall basic parameters describing cosmological first-order phase transitions. In Section \ref{sec:matching_equations}, after a brief discussion of the stationary profiles, we derive and generalize the third matching condition beyond the local thermal equilibrium approximation. We focus on the particular form of the entropy current widely used in the literature \cite{Balaji:2020yrx, Ignatius:1993qn}, however, our results are easy to adjust for alternative expressions. Then, using the new condition in Section~\ref{sec:analytical_results}, we discuss model-independent results for the wall velocity beyond the local thermal equilibrium approximation.
Section \ref{sec:simulations} presents the setup of the real-time hydrodynamic simulations used to verify the analytical solutions and their results for two representative benchmarks, comparing them with our analytical estimates. Appendix \ref{app:Wall_width} is devoted to the discussion of different approximations of the bubble-wall width, which appears in the third matching condition beyond the LTE limit.

\section{Cosmological first-order phase transition parameters}\label{sec:2}

Cosmological first-order phase transitions (FOPTs) proceed via nucleation of bubbles containing the energetically favourable phase in the background of the metastable phase. The tunneling probability per unit time and volume at temperature $T$ is given with~\cite{Coleman:1977py,Callan:1977pt,Linde:1980tt,Linde:1981zj}
\begin{equation}
\Gamma(T) = A(T)\textrm{e}^{-S}\, ,
\end{equation}
where for finite temperatures the Euclidean action $S=\frac{S_3}{T}$ and 
$A(T)=T^4\left(\frac{S_3}{2\pi T}\right)^{\frac{3}{2}}$. Nucleation temperature is defined as the value such that the probability of a true vacuum bubble forming within a horizon radius is close to unity~\cite{Ellis:2018mja}, i.e. 
\begin{equation}
N(T_n) = \int_{T_n}^{T_c} \frac{\textrm{d}T}{T} \frac{\Gamma(T)}{H(T)^4} \approx 1, 
\label{eq:nucleation}
\end{equation}
with the critical temperature $T_c$ representing the value at which both minima of the model-dependent effective potential are equally deep. Assuming that the transition is sufficiently fast, one can neglect the background evolution of the Universe assuming $H(t)\approx$ const. Then, eq. \eqref{eq:nucleation} becomes
\begin{equation}
\frac{S_3}{T_n}\approx 4\log\left(\frac{T_n}{H}\right),
\end{equation}
which at the electroweak scale gives $S_3/T_n \approx 140$~\cite{Caprini:2019egz}. 

In the standard hydrodynamic approximation~\cite{Espinosa:2010hh, Konstandin:2010dm}, it is assumed that the cosmic plasma can be modelled as the relativistic perfect fluid, with the energy-momentum tensor given by
\begin{equation}
    T^{\mu \nu}_{\textrm{fluid}} = w u^{\mu}u^{\nu} - g^{\mu\nu}p, \label{eq:energy-momentum_tensor_fluid}
\end{equation}
where $u^{\mu}$ is the four-velocity of the plasma (normalized such that $u_\mu u^\mu = 1)$, while $p$ and $w$ are respectively the pressure and the enthalpy density.
Following \cite{Ai:2021kak,Ai:2023see}, we define the transition strength as the normalized difference of the so-called pseudotrace between the symmetric and broken phase 
\begin{equation}
    \alpha_{\theta} = \frac{\Delta\theta}{3 w_s}\, , \qquad\textrm{with}\qquad \theta = e -\frac{p}{c_b^2}\, ,
\end{equation}
where $c_b$ is the speed of sound in the broken phase, while $e = w - p$ is the fluid energy density. Additionally, we use the ratio of enthalpy between broken and symmetric phases
\begin{equation}
    \Psi = \frac{w_b}{w_s}.
\end{equation}
Both $\alpha_{\theta}$ and $\Psi$ are evaluated at the nucleation temperature $T_n$.
\section{Matching conditions beyond local thermal equilibrium}\label{sec:matching_equations}
\subsection{Stationary bubble-walls}
 Conservation of the energy-momentum tensor \eqref{eq:energy-momentum_tensor_fluid}, together with the assumptions of the scale invariance and spherical symmetry of the solution leads to the following set of differential equations describing thermodynamical profiles of the plasma~\cite{Espinosa:2010hh}
\begin{equation}\label{eq:v_diff}
    \begin{aligned}
        &2\frac{v}{\xi}=\gamma^2(1-v\xi)\Big[\frac{\mu^2(\xi, v)}{c^2}-1 \Big]\partial_\xi v,\\
        &\partial_\xi w=w\Big(1+\frac{1}{c^2}\Big)\gamma^2\mu \partial_\xi v,
    \end{aligned}
\end{equation}
where the self-similar variable $\xi\equiv r/t$ is the distance from the bubble centre divided by the time elapsed since nucleation, and $c$ stands for the speed of sound. The plasma velocity $v$ is defined in the reference frame of the bubble center and 
\begin{align}
    \gamma&\equiv\frac{1}{\sqrt{1 - v^2}}, & \mu(\xi, v)&\equiv\frac{\xi-v}{1-\xi v}.
\end{align}

Boundary conditions, necessary to solve this system and obtain the profile of plasma velocity and enthalpy are derived from the conservation of the energy-momentum tensor across the wall
\begin{equation}
\label{eq:T_cons}
\partial_\mu T^{\mu\nu}_{\textrm{fluid}}=0.
\end{equation}
By integrating \eqref{eq:T_cons} over the bubble profile at steady state within the planar wall approximation, one obtains the two well-known matching conditions
\begin{align}
    &w_{+}\gamma^2_{+} v_{+}=w_{-}\gamma^2_{-} v_{-},\label{eq:match_1}\\
    &w_{+}\gamma^2_{+} v^2_{+}+p_{+}=w_{-}\gamma^2_{-} v^2_{-}+p_{-},\label{eq:match_2}
\end{align}
where all thermodynamical quantities are evaluated respectively in front of $(+)$ and behind $(-)$ the wall. These conditions are obtained in the \emph{reference frame of the bubble wall}, but we assume trivial transformation of entropy and pressure and, therefore the $w_{\pm}$ and $p_{\pm}$ can be used also in the plasma rest frame.

The resulting thermodynamical profiles constitute a family of solutions which can be parameterized with the transition strength $\alpha_{\theta}$ and the bubble-wall velocity $\xi_w$ (assuming $c^2_s = c^2_b = \frac{1}{3}$, however solutions can be easily generalized beyond the limit of the relativistic gas for the speed of sound, see \cite{Giese:2020znk, Ai:2023see}). To eliminate one of them, an additional constraint is needed. Recently it was proposed \cite{Ai:2021kak, Ai:2023see} that the conservation of entropy can be used, assuming the limit of local thermal equilibrium. The next section is devoted to our derivation of a more general form of this condition.

\subsection{Third matching condition} 

The additional relation between the thermodynamic quantities in both phases can be derived by analysing the entropy production in the steady state of the system. We assume that the only process in which the entropy is produced during FOPT is a dissipative friction-like interaction of thermal plasma with the moving wall. We focus on the regime where this interaction is local and depends on the microscopic structure of the model, plasma velocity, field profile and temperature across the wall
\begin{equation}\label{eq:Div_s}
        \partial_\mu  (u^\mu s)= f_s\,(v, \phi, T).%\frac{\eta}{T}(u^\mu\partial_\mu \phi)^2
\end{equation}
By integrating \eqref{eq:Div_s} over bubble wall \emph{in the wall frame}, one obtains
\begin{equation}
    \frac{\partial }{\partial t}\Big(\int_{M}\textrm{d}V s\,\gamma \Big)-\int_{M}\textrm{d}V\Big(\nabla\cdot (s\gamma\vec{v})\Big)=\int_{M} \textrm{d}V\, f_s(v, \phi, T), %\frac{\eta}{T}(u^\mu\partial_\mu \phi)^2,
\end{equation}
where $M$ is a thin shell with $\partial_r\phi\neq0$. Assuming a steady state, the total amount of entropy in the bubble wall does not change and hence the time derivative of the total entropy vanishes.

We use Gauss's theorem to trade volume integral over entropy current divergence for the flux through the boundaries of our region. In the planar wall limit, this condition reads
\begin{equation}
    s_{-}\gamma_{-}v_{-} - s_{+}\gamma_{+}v_{+} =\int^{l/2}_{-l/2}  \textrm{d}z\, f_s(v, \phi, T), 
\end{equation}
where $l$ is the width of shell $M$ measured as a~distance $z$ from the centre of the wall. The associated error is proportional to $L_w/r$, which is tiny in realistic situations.

This expression is identical to the familiar local thermal equilibrium condition~\cite{Ai:2021kak, Ai:2023see} after setting $f_s=0$.
The inhomogeneity given by the integral over field profile can be interpreted as the entropy production rate at the bubble-front and will be denoted with
\begin{equation}
    \Delta S = \int^{l/2}_{-l/2}\textrm{d}z\,f_s(v, \phi, T). 
\end{equation}
With the help of the definition of enthalpy $s=\frac{w}{T}$ and the matching condition \eqref{eq:match_1}, we find
\begin{equation}\label{eq:match_DSb}
    \frac{T_{+}}{T_{-}}=\frac{\gamma_{-}}{\gamma_{+}}\left(1+\frac{T_{+}\Delta S}{w_{+}\gamma_{+}v_{+}}\right)\,
\end{equation}
which is the generalised form of our third matching condition.

\subsection{Entropy production details}

Evaluation of $\Delta S$ requires the form of the entropy source $f_s$, which is predetermined by the microscopic structure of the plasma in a particular model. The calculation involves solving the associated system of Boltzmann equations, which can be approached numerically~\cite{Ekstedt:2024fyq}. The analytical approximations were derived for the SM~\cite{Moore:1995si} and the Minimal Supersymmetric Standard Model~\cite{John:2000zq} assuming a~small departure from LTE, which is valid for weak FOPTs. Here we use a simplified description, utilizing a phenomenological ansatz~\cite{Ignatius:1993qn,Kurki-Suonio:1995yaf,Balaji:2020yrx} which is also widely used in lattice studies of this system~\cite{Hindmarsh:2013xza,Hindmarsh:2015qta,Hindmarsh:2017gnf,Cutting:2019zws,Cutting:2022zgd}

\begin{equation}\label{eq:s_source}
    f_s(v, \phi, T)=\frac{\eta}{T}(u^\mu\partial_\mu \phi)^2.
\end{equation}
The effective coupling $\eta$, treated here as a free parameter, encodes the microscopic model structure. 

The scalar field derivative $\partial_\mu\phi$ is non-zero only at the bubble front, and thus we can estimate $\Delta S$ by setting $1/T$ and $u^{\mu}u^{\nu}$ to their values in front of the bubble wall\footnote{From the field profile alone, there is no fundamental reason why $\Delta S$ should be estimated using only the quantities in front of the bubble. However, when the transition is relatively weak, as in the cases considered here, the differences between the values in front of and just behind the bubble wall are small, so more complex expressions do not result in any significant numerical difference.} as
\begin{equation}\label{eq:Delta_S1}
    \Delta S\approx\frac{\eta}{T_{+}} u^\mu_{+} u^\nu_{+} \int^{l/2}_{-l/2} \textrm{d}z\,  \partial_\mu \phi \partial_\nu \phi.
\end{equation}

To obtain analytically tractable expressions, we use the Tanh-ansatz:
\begin{equation}\label{eq:phi_prof}
    \phi(z,t)=\frac{\upsilon_0}{2}\left[1-\tanh\left(\frac{z}{L_w}\right)\right],
\end{equation}
suitable for most physical cases~\cite{Moore:1995si}. Here $\upsilon_0$ is the scalar vacuum expectation value (VEV) at the broken phase, and $L_w$ is the field profile width. With the field profile \eqref{eq:phi_prof},
the integral \eqref{eq:Delta_S1} yields
\begin{equation}\label{eq:Delta_S2}
   \frac{\eta(v_{+}\gamma_{+})^2}{T_+} \int^{\infty}_{-\infty} \textrm{d}z  (\partial_z \phi)^2 = \frac{\eta}{3}\frac{\upsilon_0^2}{L_w}\frac{\gamma_{+}^2v_{+}^2}{T_+}.
\end{equation}
where the integral was computed in the limit $l  \rightarrow \infty$, since the derivative of the Tanh-ansatz decreases fast away from $M$, and thus the integration may be extended to the whole real domain. 

After inserting \eqref{eq:Delta_S2} into \eqref{eq:match_DSb}, the third matching condition becomes
\begin{equation}\label{eq:match_3}
    \frac{T_{+}}{T_{-}}=\frac{\gamma_{-}}{\gamma_{+}}\,\left(1+\tilde{\eta}\,\gamma_{+}\,v_{+}\right),
\end{equation}
where $\tilde{\eta}$ is the rescaled friction parameter, defined as
\begin{equation}
\tilde{\eta}\equiv \chi\,\eta,\qquad \chi\equiv\frac{ \upsilon_0^2}{3 w_{+} L_w} . 
\label{eq:eta_tilde}
\end{equation}

As long as the transition is not too strong $(\alpha_{\theta} \lesssim 0.1)$, $w_{+}$ can be well-approximated by the enthalpy at the symmetric phase $w_s$ and hence, the scaling coefficient $\chi$ is approximately friction-independent. Note that in the limit $\tilde{\eta}\rightarrow 0$, one retrieves the matching condition for the LTE regime~\cite{Ai:2021kak,Ai:2023see}.

It has been recently proven \cite{Ai:2024uyw, Krajewski:2024xuz} that the entropy production rate is maximized in the ballistic regime, where the mean free path of the particles is much larger than the bubble wall width $L_w$ but smaller than the size of the plasma shell. The maximal entropy production rate depends on the ratio of the mass gained by the particles predominantly responsible for friction to the plasma temperature. A universal, model-independent bound is obtained when this ratio is large. Assuming simplified Bag equations of state~\cite{Chodos:1974je}, it takes the form~\cite{Ai:2024uyw}
\begin{equation}
    \frac{\Delta S}{s_{+}\gamma_{+}v_{+}}<\left(\Psi \frac{v_{-}\gamma_{+}^2}{v_{+}\gamma_{-}^2}\right)^{\frac{1}{4}}-1.
\end{equation}
This condition translates into an upper limit on the rescaled friction parameter:
\begin{equation}
        \tilde{\eta}_{\rm max}=\frac{1}{\gamma_+ v_+}\left[\left(\Psi \frac{\gamma_{+}^2v_{-}}{\gamma_{-}^2v_{+}}\right)^{\frac{1}{4}}-1\right],
        \label{ballistic_limit}
\end{equation}
which defines the range of applicability of the methods used in this work. Near this boundary, the system deviates significantly from the LTE approximation, making the standard hydrodynamic treatment unreliable.

In fact, the entropy production ansatz given in Eq.~\eqref{eq:s_source} is rather simple, which limits the scope of its applicability. It assumes a local form of friction, valid only in the regime of strong interactions and small deviations from equilibrium distribution functions of plasma species. It is not expected to fully reproduce the non-local results obtained from solving the Boltzmann equation, especially near the ballistic regime. Therefore, we recommend using a conservative applicability range, restricted to $\tilde{\eta} \ll \tilde{\eta}_{\textrm{max}}$.

\section{Analytical results}\label{sec:analytical_results} 
We solve the system~\eqref{eq:v_diff} utilizing new matching condition~\eqref{eq:match_3} together with two standard ones~\eqref{eq:match_1} and~\eqref{eq:match_2}. The problem is handled with the modified \texttt{Python} code from~\cite{Ai:2023see}, publicly available at~\cite{Repo:2024in}.

\begin{figure}[!t]
    \centering
    \includegraphics[width=0.49\linewidth]{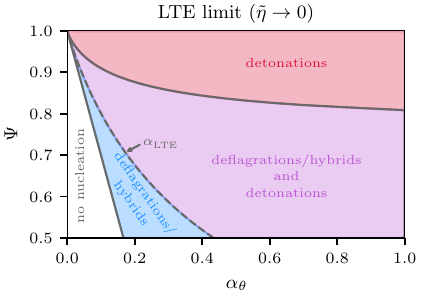}
    \includegraphics[width=0.49\linewidth]{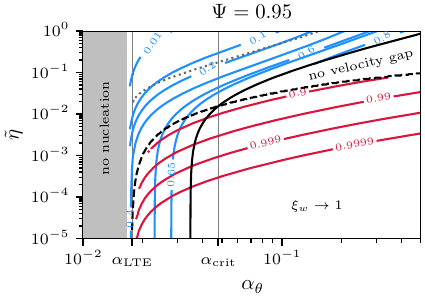}
    \caption{ \textit{
    \textbf{Left panel:} LTE limit of the hydrodynamic solution. The blue region corresponds to the case when only deflagration/hybrid branch exists for $\tilde{\eta}\rightarrow 0$. The purple region denotes the situation where both deflagration/hybrid and detonation branches coexist, however, the detonation solution is typically realized in lattice simulations. The red region corresponds to the situation with no deflagration/hybrid solution (runaway scenario). \\
    \textbf{Right panel:} contour plot of the bubble wall velocity in $(\alpha_\theta$, $\tilde{\eta})$ plane. The red contours represent detonation solutions, while deflagration/hybrid profiles are colored in blue. The solid black line marks the contour below which there is no stable deflagration or hybrid, while the black dashed line limits the region where detonation solutions exist. In the gray shaded region, nucleation is forbidden. The gray dotted line illustrates the upper bound on $\tilde{\eta}$, coming from the ballistic limit, given with Eq. \eqref{ballistic_limit}.}}
    \label{nonLTE_predictions}
\end{figure}

The left panel of Fig.~\ref{nonLTE_predictions} shows the limiting case with $\tilde{\eta}\to 0$ corresponding to the LTE approximation. Different types of stationary solutions are shown on the $(\alpha_{\theta}, \Psi)$ plane, where $\Psi = w_b / w_s$ denotes the ratio of enthalpy between symmetry-breaking and symmetry-preserving phases. In most of the parameter space, we find stable detonation solutions, which for $\tilde{\eta} = 0$ become runaways.~\footnote{Reference~\cite{Ai:2021kak} also finds detonation solutions for some benchmarks in LTE but correctly classifies them as unphysical. Here, we found a different branch of stable detonation solutions, which do not exist in LTE. For all solutions on this new branch, we find $v_w\rightarrow 1$ in the limit $\eta\rightarrow 0$ which corresponds to LTE.} Dynamic lattice simulations discussed in the next section indicate that for realistic initial conditions, the run-away branch is preferred even if there exist other stationary solutions. Thus, the LTE deflagration/hybrid solutions found in~\cite{Ai:2021kak} are realized in the blue region, where no run-away solutions exist and are disfavoured in the purple one where detonations are possible. 

The map of the existing stationary states in $(\alpha_\theta$, $\tilde{\eta})$ plane is presented in the right panel of Fig.~\ref{nonLTE_predictions}.  
The solid black line corresponds to the Jouguet velocity $c_J$ \cite{Ai:2023see}
\begin{equation}
c_J\equiv c_b\left(\frac{1+\sqrt{3\alpha_{\theta}(1-c_b^2+3c_b^2\alpha_{\theta})}}{1+3c_b^2\alpha_{\theta}} \right),
\end{equation}
indicating the fastest deflagrations/hybrids, while the dashed black line shows where the slowest detonation solutions lie.

The bottom of the plot corresponds to the LTE limit ($\tilde{\eta}\rightarrow 0$). For relatively strong transitions, we found only ultrarelativistic detonations with $v_w\to 1$, while for moderate and weak transition strength we find also deflagrations and hybrids. As the friction parameter grows, the wall velocity decreases, and for detonation solutions, it approaches the lower velocity limit, denoted with the black dashed line. When friction is even larger, we find only the deflagration/hybrid branch, and the wall velocity continues to decrease with the increasing friction parameter.

For latent heat below the threshold value $\alpha_{\textrm{crit}}$, the velocity of the slowest detonation is above $c_J$, while the corresponding deflagrations/hybrids crossing the dashed line are slower than $c_J$. Therefore, we find a gap in allowed velocities discussed in~\cite{Krajewski:2023clt}. For stronger transitions with \mbox{$\alpha_{\theta}>\alpha_{\textrm{crit}}$}, all the velocities are kinematically allowed, yielding a continuous transition between the hybrids and detonations.

If the transition is sufficiently weak ($\alpha_{\theta}<\alpha_{\textrm{LTE}}$), only deflagration/hybrid solutions exist for any value of $\tilde{\eta}$. 
In this region, for $\tilde{\eta}\rightarrow 0$, pure hydrodynamic obstruction can stop the wall acceleration~\cite{Konstandin:2010dm}, and thus, the LTE estimate of the bubble-wall velocity~\cite{Ai:2023see} is realized. The value of $\alpha_{\textrm{LTE}}$ is determined numerically as the asymptotic behavior of the black dashed curve for $\tilde{\eta}\to0$. It corresponds to the dashed line between the blue and purple regions on the left panel of Fig. \ref{nonLTE_predictions}. 

The existence of the detonation branch provides a straightforward criterion for determining whether the bubble wall reaches the deflagration/hybrid steady state or continues to accelerate. Specifically, for benchmarks where a detonation solution does not exist, regardless of the $\tilde{\eta}$ value, walls cannot achieve ultrarelativistic velocities. In such cases, the only possible solution is a deflagration or a hybrid. Consequently, this yields a simple condition for assessing the validity of LTE predictions regarding bubble wall velocities. It is important to note that this condition is derived in the limit $\tilde{\eta} \rightarrow 0$ and is, therefore, independent of our specific choice for the entropy source \eqref{eq:s_source}.

\section{Hydrodynamic simulations}\label{sec:simulations}
To verify the analytical predictions, we performed real-time simulations of the bubble growth. Similarly as in \cite{Ignatius:1993qn, Kurki-Suonio:1995yaf, Hindmarsh:2013xza, Hindmarsh:2015qta, Hindmarsh:2017gnf,Cutting:2019zws, Cutting:2022zgd}, we describe the plasma as a perfect fluid with temperature $T$, characterized by the internal energy density $e$, pressure $p$ and enthalpy density $w$. As the effective potential $V_{\textrm{eff}}(\phi, T)$ can be interpreted as the free energy density $\mathcal{F}$ of the system, we define the equation of state as 
\begin{align}
p(\phi, T) &= - V_{\textrm{eff}}(\phi,T),\label{eq:pressure}\\
e(\phi, T) &= V_{\textrm{eff}}(h, s, T) - T\frac{\partial V_{\textrm{eff}}(\phi, T)}{\partial T}, \label{eq:energy}\\
w(\phi, T) &= -T\frac{\partial V_{\textrm{eff}}(\phi, T)}{\partial T}\label{eq:entalpy}.
\end{align}
The energy-momentum tensor of such a system is given by
\begin{equation}
    T^{\mu \nu} = T^{\mu \nu}_{\textrm{field}} + T^{\mu\nu}_{\textrm{fluid}},
\end{equation}
where $T^{\mu\nu}_{\textrm{fluid}}$ is defined as in \eqref{eq:energy-momentum_tensor_fluid}, while for the scalar field~\footnote{Note that in our convention the vacuum energy is a part of the fluid energy-momentum tensor and is not present in the field's part.}
\begin{equation}
    T^{\mu \nu}_{\textrm{field}} = \partial^\mu \phi \partial^\nu \phi  - g^{\mu \nu}\left( \frac{1}{2} \partial_\alpha \phi \partial^\alpha \phi \right).
\end{equation}

The total energy-momentum tensor of the system is conserved $(\nabla_\mu T^{\mu \nu} = 0 )$, however, both contributions are not conserved separately
\begin{equation}
    \nabla_\mu T^{\mu \nu}_{\textrm{field}} = - \nabla_\mu T^{\mu \nu}_{\textrm{fluid}} = \frac{\partial V_{\textrm{eff}}}{\partial \phi} \partial^\nu \phi + \eta u^\mu \partial_\mu \phi \partial^\nu \phi\, ,\label{eq:energy_momentum_conservation}
\end{equation}
where the right-hand side can be interpreted as the interaction between the plasma and the field which consists of the equilibrium part determined by $V_{\textrm{eff}}(\phi, T)$ and the dissipative friction, parametrized by $\eta$, that is responsible for the entropy production at the bubble front, as in \eqref{eq:s_source}.

Assuming spherical symmetry of the problem and introducing variables  $Z:=w\gamma^2v$ and $\tau:=w\gamma^2 - p$, we get in the plasma rest frame
\begin{align}
&\begin{aligned}\label{eq:EOM_field}
-\partial_t^2\phi + \frac{1}{r^2}\partial_r(r^2\partial_r\phi) =
\frac{\partial V_{\textrm{eff}}}{\partial\phi} +  \eta\gamma(\partial_t\phi + v\partial_r\phi)\,
\end{aligned}\\
&\begin{aligned}\label{eq:EOM_fluid1}
\partial_t \tau + \frac{1}{r^2} \partial_r (r^2 (\tau + p) v)
= \frac{\partial V_{\textrm{eff}}}{\partial \phi} \partial_t \phi + \eta \gamma (\partial_t \phi + v \partial_r \phi) \partial_t \phi\, 
\end{aligned}\\
&\begin{aligned}\label{eq:EOM_fluid2}
\partial_t Z + \frac{1}{r^2} \partial_r \left(r^2 Zv \right) + \partial_r p = -\frac{\partial V_{\textrm{eff}}}{\partial \phi} \partial_r \phi - \eta \gamma (\partial_t \phi + v \partial_r \phi) \partial_r \phi.\\
\end{aligned}
\end{align}
This, together with the equations of state \eqref{eq:pressure}--\eqref{eq:entalpy}, defines the dynamics of the system which we solve numerically on a lattice.

As a practical example, we use the popular scalar singlet model (xSM)~\cite{McDonald:1993ey,Espinosa:1993bs,Espinosa:2007qk, Profumo:2007wc,Espinosa:2011ax,Barger:2011vm,Cline:2012hg, Alanne:2014bra, Curtin:2014jma,Vaskonen:2016yiu,Kurup:2017dzf,Beniwal:2017eik, Beniwal:2018hyi,Niemi:2021qvp,Ellis:2022lft,Lewicki:2024xan, Gould:2024jjt,Ramsey-Musolf:2024ykk}, although features we discuss are general and would carry over to many other extensions of the SM. We generalized the equation \eqref{eq:EOM_field}--\eqref{eq:EOM_fluid2} for two fields $\phi = (h,s)$ (see \cite{Krajewski:2024gma}). Similarly as in \cite{Cline:2021iff,Laurent:2022jrs}, we assume that the non-equilibrium friction is produced only through the coupling to the Higgs field, therefore from now on we assume $\eta_h=\eta$ and $\eta_s = 0$.  For the details of the numerical treatment see Appendix A of~\cite{Krajewski:2024gma}. 

We compare our analytical predictions for wall velocity, derived using the generalized third matching equation \eqref{eq:match_3}, with the results of dynamic lattice simulations for two representative benchmark points (see Table \ref{tab:benchmarks}). 

\begin{table}[t]
    \centering
    \renewcommand{\arraystretch}{1.1}
    \begin{tabular}{|c|c|c|c|c|c|c|}
    \cline{2-7}
        \multicolumn{1}{c|}{} & $\alpha_{\theta}$ & $\Psi_n$ &  $w_s(T_n) \; [\text{GeV}^4]$ & $T_n \; [\text{GeV}]$ & $h_0 \; [\text{GeV}]$ & $L^{\text{app}}_w \; [\text{GeV}^{-1}]$\\
    \cline{2-7}
    \hline
        \textbf{Benchmark $1$} & $0.0072$  & $0.979$ 
        & $3.87\times 10^{9}$ & $96.8$ & $172$ & $0.181$\\
        \textbf{Benchmark $2$} & $0.0103$   & $0.972$ & $2.94\times 10^9$ & $90.6$ & $182$& $0.079$\\

    \hline
    \end{tabular}
    \caption{\textit{Benchmark realisations of the xSM model: \textbf{Benchmark $1$} is an example of the model which evolves towards stationary subsonic deflagration in the limit of $\eta\rightarrow 0$.  In the same limit \textbf{Benchmark $2$} evolves as the ultra-relativistic detonation, although the analytical deflagration solution, predicted by the LTE equations, exists. The scan with respect to the friction parameter $\eta$ is presented on Fig~\ref{fig:v_w-eta}}}
    \label{tab:benchmarks}
\end{table}

In order to translate the friction parameter $\eta$ used in the hydrodynamic simulation into the normalized friction parameter $\tilde{\eta}$, one needs to compute the width of the wall $L_w$ for the stationary field profile. For this purpose, one can use the analytical approximation 
\begin{equation}\label{eq:L_wTc}
    L_w\approx\frac{ h_0}{\sqrt{8 V_b|_{T_c}}}
\end{equation}
which depends on the Higgs VEV $h_0$ and the potential barrier between the supercooled and true vacua on the tunneling path, evaluated at the \emph{critical} temperature. As we show in Appendix~\ref{app:Wall_width}, this simple expression is enough to give an accurate approximation of the wall velocity.

\begin{figure}[t]
    \centering
    \includegraphics[scale=1]{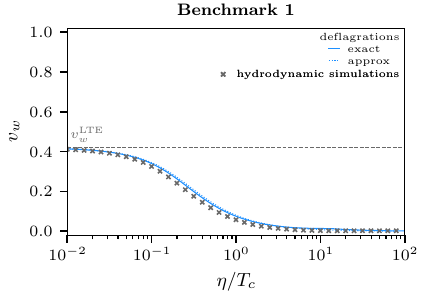}
    \includegraphics[scale=1]{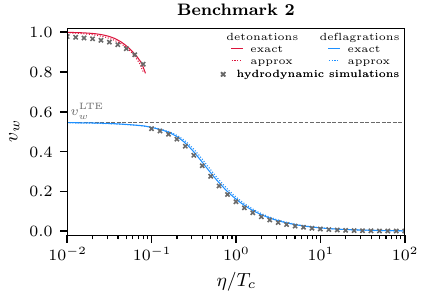}
    \caption{\textit{The bubble wall velocity $v_w$ in the stationary state plotted against the effective friction parameter $\eta$ for two benchmarks (see Table \ref{tab:benchmarks}). Red and blue curves represent the analytical results with the third matching condition beyond LTE given with eq. \eqref{eq:match_3}. Solid and dotted lines denote different methods of bubble wall width evaluation: the exact value measured from the simulation, and the analytical approximation \eqref{eq:L_wTc}.  Gray dashed lines mark the standard LTE analytical results (see \cite{Ai:2023see}), while grey crosses represent the results of real-time lattice simulations with the full equations of motion of the system.
    }}
    \label{fig:v_w-eta}
\end{figure}

Fig. \ref{fig:v_w-eta} shows our predictions for the terminal wall velocity with the friction coupling $\eta$ treated as a free parameter. The left panel represents the benchmark with $\alpha_{\theta} < \alpha_{\textrm{LTE}}$, where in the LTE limit hydrodynamic obstruction stops the acceleration and the stationary state is achieved. The right panel corresponds to the benchmark with $\alpha_{\theta}>\alpha_{\textrm{LTE}}$, for which two distinct steady-states are kinematically allowed. Grey crosses show the values computed directly from simulations.
Our steady-state asymptotic predictions, determined with extended matching conditions at different $\eta$ are plotted with colored lines. The deflagration/hybrid branch of solutions is denoted with blue, while the detonation branch is in red, similar to Fig.~\ref{nonLTE_predictions}. The solid lines show our analytical estimates discussed in Sec.~\ref{sec:matching_equations} with the wall width from the simulation used in Eq.~\eqref{eq:eta_tilde}, while to compute the dashed lines we use the approximation for the width from Eq.~\eqref{eq:L_wTc} instead. We can see that the uncertainty associated with that simple estimate is very small.
The horizontal gray dashed line shows the velocity in the LTE limit. 

When there is more than one stable asymptotic solution for the PDE, it is the initial state that determines which of the steady states is realized. Starting with the nucleating bubble, our simulations show it is the faster solution that is strongly preferred. The reason is that the fluid shell is still forming when the wall accelerates past the slower solution's velocity and the friction profile at that time does not resemble the asymptotic one which analytical methods assume~\footnote{We discussed the same issue in LTE with $\eta=0$ in ref.~\cite{Krajewski:2024gma} where walls would accelerate past the deflagration/hybrid LTE solutions and run away.}.

\begin{figure}[t]
    \centering
    \includegraphics[scale=1]{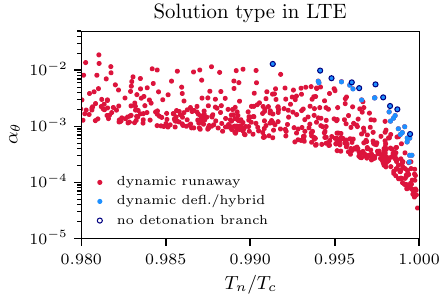}
    \caption{\textit{Solution types in LTE regime predicted by the lattice simulations for a large set of xSM benchmarks (see \cite{Krajewski:2024gma}). In the plotted region, the stationary LTE solution is only realized dynamically at benchmarks plotted in blue, while for the ones plotted in red, the wall runs away. Dark blue rim marks points for which analytic computation does not find detonation branch and, therefore, the system is expected to reach deflagration/hybrid solution. This analytical prediction is confirmed by lattice simulations that converge to deflagration/hybrid solutions for all rimmed benchmarks.}}
    \label{fig:sol_type}
\end{figure}

To verify the accuracy of our selection rule, we applied it to the sample studied in \cite{Krajewski:2024gma}. As shown in Fig. \ref{fig:sol_type}, all benchmarks for which our method does not predict the existence of a detonation branch dynamically evolve into stationary deflagrations or hybrid solutions. This suggests that the absence of a detonation branch serves as a sufficient condition for steady-state expansion.

The remaining deflagration/hybrid solutions may arise due to small numerical friction present in the simulations due to unavoidable imperfections of the numerical scheme. As is visible on the right panel of Fig. \ref{nonLTE_predictions}, for fine-tuned values of strength of the transition $\alpha$ close to $\alpha_{\textrm{LTE}}$, detonations exist only for extremely small values of $\tilde{\eta}$ which can be saturated by the numerical friction in our simulations. Another source of uncertainty arises from the approximations used to derive the analytical relation between friction and velocity. We cannot also exclude the possibility that the deflagrations/hybrids observed in numerical simulations, which are not predicted by our condition, speed up slowly enough to form heated fluid shells that inhibit further acceleration before reaching Jouguet velocity and finally reach the stationary state with the finite terminal velocity. Such a situation is, however, highly fine-tuned and in the vast majority of the parameter space, bubbles expand as runaways in LTE approximation when the detonation branch is present, as we can see in Fig.~\ref{fig:sol_type}.

\section{Conclusions}

In this work, we investigate the possible hydrodynamic solutions for a steady-state bubble wall in cosmological phase transitions. We extend the analytical computation of the wall velocity, showing that the rate of entropy production fully determines this parameter. We perform an analytical computation involving matching conditions across the wall and extend them with a new equation describing the entropy production. This allows us to predict the wall velocity and find a branch of stable detonation solutions which is not present in LTE regime (where $\Delta S = 0$). We thus show that due to the entropy production, two stable solutions with different wall velocities typically coexist. This result neatly explains all the phenomena observed so far in lattice simulations. The code we provide~\cite{Repo:2024in} allows one to easily map all possible solutions in the entire parameter space of a model without relying on expensive simulations.

Next, we use real-time lattice simulations initiated with nucleating bubble solutions to verify our results. We find our analytical solutions to match the lattice results very well. We observe that above a certain critical value for the phase transition strength, there is no gap between the velocities of the two branches of solutions and, in principle, all the profiles are kinematically allowed. Below that critical value, the deflagration/hybrid solutions often coexist with the new detonation branch, however, there is a velocity gap between the two. Using our real-time simulations, we demonstrate that the faster detonation solution is predominantly realized in such cases. The reason behind this is that in analytical methods, we assume a fully developed steady-state thermodynamic profile for all solutions. In simulations, we see that the plasma shell is typically still forming and exerts much smaller friction as the wall accelerates past the slower solution. This allows us to predict the character of solutions in the entire parameter space of models, providing a simple criterion determining whether the bubble wall runaways.

The main limitations of the method lie in the uncertainty regarding the range of applicability of the ansatz assumed for the entropy source and the difficulty of determining the exact value of the $\eta$ parameter for a given model, which is essential for assessing how far the system is from local thermal equilibrium. These issues can, in principle, be addressed by solving the Boltzmann equation across the wall and we leave it for future work.

\subsubsection*{Acknowledgements}
We would like to acknowledge Bogumiła Świeżewska, Jos\'e Miguel No and Marcin Badziak for the discussion on the draft. ML and MZ were supported by the Polish National Agency for Academic Exchange within Polish Returns Programme under agreement PPN/PPO/2020/ 1/00013/U/00001 and the Polish National Science Center grant 2023/50/E/ST2/00177. IN was partially supported by the National Science Center, Poland, under research grant no. 2020/38/E/ST2/00243. TK was supported by Polish National Science Centre grant 2019/33/B/ST9/01564. MZ and IN were also supported with the IDUB Early Universe scholarships.

%%%%%%%%%%%%%%%%%%%%%%%%%%%%%%%%%%%%%%%%%%%%%%%%%%%%%%%%%%%%%%%%%%%%%%%%%%%
\appendix
\section{Field profile widths}
\label{app:Wall_width}
In this Appendix, we discuss steady-state field profiles across the wall. We begin from the scalar field equation of motion (EOM) written in the planar wall limit  
\begin{equation}\label{eq:phi}
    -\partial^2_t\phi+ \partial^2_z\phi-\frac{\partial V_{\text{eff}}}{\partial \phi}=\eta \gamma (\partial_t\phi+ v\partial_z \phi)
\end{equation}
were $\phi$ could be a single field or a vector of dynamical scalar fields that tunnel through the potential barrier. We neglected the effects of the wall curvature since the associated error is proportional to the inverse of the bubble radius and should be negligible in late stages of its evolution.

\begin{figure*}[t!]
    \centering
    \includegraphics[scale=1]{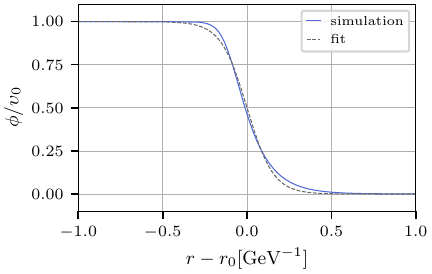}
    \includegraphics[scale=1]{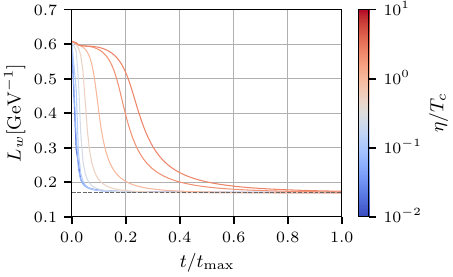}
    \caption{\it{Left panel: example of the field profile from the real-time simulation at its final time $t=t_{\textrm{max}}$ compared with the fit given with eq. \eqref{eq:ansatz}.  Right panel: time evolution of $L_w$ for \textbf{Benchmark 1} (see Table I in the main text). Different colours represent different values of the effective friction parameter $\eta$. The asymptotic value $L_w^{\textrm{LTE}}$ is achieved in the whole spectrum of $\eta$ values.}}
    \label{fig:Lw_fits}
\end{figure*}

To get the equation for a stationary profile in the wall frame, we set all time derivatives of $\phi$ to zero. The lattice simulation results indicate that for weak and moderate transition strength, this profile weakly depends on $\eta$ (see the right panel in Fig. \ref{fig:Lw_fits}). Therefore, we can safely neglect the friction term in Eq. \eqref{eq:phi}, at least for deflagrations and hybrids
 \begin{equation}\label{eq:phi_inf}
    \partial^2_z\phi-\frac{\partial V_{\text{eff}}}{\partial \phi}=0.
\end{equation}

The resulting equation is equivalent to the classical equation of motion of a particle with ``position''  $\phi$, ``time'' $z$ and ``potential'' $-V_{\text{eff}}(\phi,T(z))$.  The potential depends on the ``time'' $z$ through the temperature, whose profile across the rarefaction front cannot be found without solving the steady-state equations of motion for the plasma.

 To simplify the problem, we focus on transitions with $T_n\sim T_c$ where the temperature does not change much across the wall. As an example, this is always true for the region of xSM parameter space where deflagrations/hybrids are dynamically realized in the LTE limit~\cite{Krajewski:2024gma}. A precise estimate of the field profile is then obtained by solving~\eqref{eq:phi_inf} at the critical temperature
 \begin{equation}\label{eq:phi_inf_Tc}
    \partial^2_z\phi=\frac{\partial V_{\text{eff}}(\phi, T=T_c)}{\partial \phi}.
\end{equation}

The potential in \eqref{eq:phi_inf_Tc} does not explicitly depend on ``time'' $z$, thus the equation has energy-like integral of motion
\begin{equation}\label{eq:E_kin}
    \frac{(\partial_z \phi)^2}{2}=V_{\text{eff}}(\phi)-V_{\text{eff}}(\left. \phi \right|_{\pm \infty}).
\end{equation}
The equation for the profile width is obtained by substituting into it hyperbolical tangent ansatz
\begin{equation}\label{eq:ansatz}
    \phi_i=\frac{ \upsilon_i }{2}\Big[1\pm \tanh\big(\frac{z-\delta_i}{L_i}\big)\Big]
\end{equation}
with some specific width $L_i$ and displacement $\delta_i$ from the center of the wall \footnote{For a sufficiently long evolution of the bubble, its wall becomes arbitrarily thin compared to the plasma profile. In this limit the notion of the bubble wall center is ambiguous. Therefore, one can pick any field profile to be ``central'' and offset its $\delta$ from other profile displacements.}. For a single-field problem, this yields the well-known expression~\cite{Huber:2013kj}
\begin{equation}\label{eq:L_wHubert}
    L_w=\frac{\upsilon_0}{\sqrt{8 V_b}},
\end{equation}
where $\upsilon_0$ is the scalar VEV at broken vacuum, while $V_b$ is the barrier height. When there are many dynamical fields whose profiles do not overlap, one can assume, that at the center of each field profile, its ``kinetic'' term  $(\partial_z \phi_i)^2/2$ dominates other derivative terms in \eqref{eq:E_kin}.

The multi-field variant of the approximation \eqref{eq:L_wHubert} is very precise, provided $V_b$ is evaluated \emph{on the tunneling path}~\footnote{The tunneling path in the vicinity of the critical temperature can be easily estimated with the routine provided by \texttt{CosmoTransitions} package \cite{Wainwright:2011kj}.} in the field space. Alternatively, one can use phenomenological approximation that does not require computation of the tunneling path
 \begin{equation}\label{eq:L_wTn}
    L_w\approx\frac{\upsilon_0}{\sqrt{2 \Delta V_{\text{eff}}}}.
\end{equation}

In the xSM precise values of the wall width can be obtained by fitting hyperbolic tangent as given by eq. \eqref{eq:phi_inf} to the field profile obtained in the lattice simulations. An example of such a fit is shown in the left panel of Fig. \ref{fig:Lw_fits}, while in the right panel, we present the time evolution of the fitted wall width towards the stationary limit. Crucially, within the deflagration/hybrid branch, the asymptotic value of $L_w$ does not depend on the effective friction $\eta$, which justifies neglecting the friction term in~\eqref{eq:phi}. We find that regardless of the method used for $L_w$ computation, our asymptotic approach is surprisingly accurate.

\begin{figure}[t]
    \centering
    \includegraphics[scale=0.94]{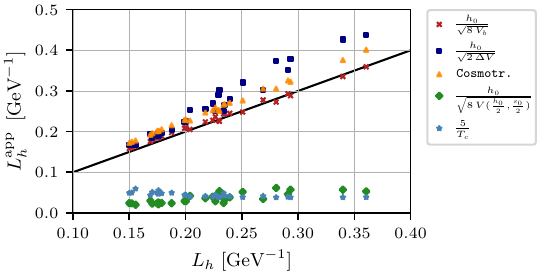}
    \caption{\textit{Different approximations for the Higgs field profile width at the steady state $L^{\text{app}}_h$, plotted against values obtained in lattice simulations $L_h$.}}
    \label{fig:Lw_approx}
\end{figure}

The fitted value of $L_w$ is used to asses analytical expressions \eqref{eq:L_wHubert} and \eqref{eq:L_wTn} as well as other estimates proposed in the literature~\cite{Huber:2013kj,Vaskonen:2016yiu, Laurent:2020gpg, Lewicki:2021pgr}. The resulting comparison is shown in Fig.~\ref{fig:Lw_approx}. 
While all the analysed methods correctly predict the order of magnitude of the Higgs profile width $L_h$, the exact values often differ significantly from lattice results. The most precise predictions, within $6\%$ from the lattice, are obtained with \eqref{eq:L_wHubert} where $V_b$ was evaluated on the tunnelling path. The naive estimate \eqref{eq:L_wTn} is also surprisingly accurate for smaller profile widths, but for larger $L_h$ it overshoots the lattice results by up to $35\%$. 
In our comparison, we also include the build-in \texttt{CosmoTransitions}~\cite{Wainwright:2011kj} routine for finding the profile of the wall moving through the plasma~\footnote{The routine numerically solves the equation \eqref{eq:phi_inf} at the nucleation temperature by adjusting $\eta$ so that the field, rolling down from the global maximum of $-V_{\rm eff}$ (true vacuum), stops exactly at the local maximum (false vacuum).}  (not to be confused with the routine computing the critical profile). This method also overshoots the lattice values, but the results are within $15\%$ from the full solutions of the EOM. The simple estimate assuming a symmetric barrier 
\begin{equation*}
    L_h\approx\frac{h_0}{\sqrt{8  V_{\text{eff}}(\frac{h_0}{2},\frac{s_0}{2})}},
\end{equation*}
predicts $L_h$ to be $5$ to $8$ times smaller than the corresponding profile widths obtained on the lattice. The approximation based on the critical temperature: 
\begin{equation*}
    L_h\approx\frac{5}{T_c},
\end{equation*}
also significantly underestimates the widths of the xSM Higgs profiles.
 
\bibliographystyle{JHEP}
\bibliography{main} 
\end{document}